\documentclass[aps,prl,twocolumn,preprintnumbers,groupedaddress]{revtex4}

\usepackage{todonotes}

\usepackage{amssymb}
\usepackage{epsfig} 
\usepackage{amsmath,color} 
\usepackage{graphicx} 
\usepackage{ulem} 
\usepackage{multirow}
\usepackage{tabularx}
\usepackage{slashed}

\usepackage{wrapfig}

\usepackage{tikz}
\usetikzlibrary{positioning,arrows}
\usetikzlibrary{decorations.pathmorphing}
\usetikzlibrary{decorations.markings}

\newcommand{\beq}{\begin{equation}}
\newcommand{\eeq}{\end{equation}}
\newcommand{\bea}{\begin{eqnarray}}
\newcommand{\eea}{\end{eqnarray}}
\newcommand{\ba}{\begin{array}}
\newcommand{\ea}{\end{array}}
\newcommand{\bi}{\begin{itemize}}
\newcommand{\ei}{\end{itemize}}
\newcommand{\bn}{\begin{enumerate}}
\newcommand{\en}{\end{enumerate}}
\newcommand{\bc}{\begin{center}}
\newcommand{\ec}{\end{center}}
\renewcommand{\l}{\left}
\renewcommand{\r}{\right}

\newcommand{\eq}[1]{Eq.~(\ref{#1})}
\newcommand{\eqs}[2]{Eqs.~(\ref{#1}) and (\ref{#2})}

\begin{document}

\preprint{FTUV-17-03-17}
\preprint{IFIC/17-19}

\title{On large lepton number asymmetries of the Universe}


\author{Gabriela Barenboim$^1$}
\email[]{Gabriela.Barenboim@uv.es}


\author{Wan-Il Park$^2$}
\email[]{wipark@jbnu.ac.kr}
\affiliation{$^1$ Departament de F\'isica Te\`orica and IFIC, Universitat de Val\`encia-CSIC, E-46100, Burjassot, Spain}

\affiliation{$^2$ Department of Science Education (Physics), Chonbuk National University, Jeonju 561-756, Korea}


\date{\today}

\begin{abstract}
A large lepton number asymmetry of $\mathcal{O}(0.1-1)$ at present universe might not only be allowed but also necessary for consistency among cosmological data.
We show that, if a sizeable lepton number asymmetry were produced before the electroweak phase transition,  the requirement for not producing too much baryon number asymmetry through sphalerons processes,  forces the high scale lepton number asymmetry to be larger than about $30$. Therefore a mild entropy release causing $\mathcal{O}(10-100)$ suppression of pre-existing particle density   should take place, when the background temperature of the universe is around $T = \mathcal{O}(10^{-2} - 10^2) {\rm GeV}$ for a large but experimentally consistent asymmetry to be present today.   
We also show that such a mild entropy production can be obtained by the late-time decays of the saxion, constraining the parameters of the Peccei-Quinn sector such as the mass and the vacuum expectation value of the saxion field to be $m_\phi \gtrsim \mathcal{O}(10) {\rm TeV}$ and $\phi_0 \gtrsim \mathcal{O}(10^{14}) {\rm GeV}$, respectively.
\end{abstract}


\maketitle


\section{Introduction}

Extensive analysis of Big Bang Nucleosynthesis (BBN) and cosmic microwave background (CMB) data showed that the baryon number asymmetry ($B$) of the Universe should be $\mathcal{O}(10^{-10})$ \cite{Cyburt:2015mya,Ade:2015xua}.
On the other hand, BBN and CMB data still allow $\mathcal{O}(1)$ lepton number asymmetries of muon- and tau-neutrinos ($L_{\mu,\tau}$) \cite{Barenboim:2016shh}, even though the asymmetry of electron-neutrinos is tightly constrained to be $|L_e| \lesssim \mathcal{O}(10^{-3})$ by BBN \cite{Mangano:2011ip}.
Such large asymmetries might provide a better fit to various astrophysical and cosmological data \cite{Barenboim:2016lxv}.

A large lepton number asymmetry ($L_\alpha \equiv \Delta n_\alpha /n_\gamma$ with $\alpha$, $\Delta n_\alpha$, and $n_\gamma$ being the neutrino flavor, the number density difference of $\nu_\alpha$ and $\bar{\nu}_\alpha$, and the photon number density, respectively) can be generated before or after the electroweak phase transition (EWPT).
If it were produced after  (at temperatures lower than) the EWPT, the anomalous electroweak processes (sphalerons) \cite{Kuzmin:1985mm,Harvey:1990qw} which effectively could transform such a lepton number asymmetry into a baryon number asymmetry (or vice versa) do not take place, i.e., a pre-existing baryon number asymmetry would not be affected by the aforesaid large lepton number asymmetry.
However, if $L_\alpha$ were produced before (at temperatures higher than) the EWPT, the danger of an over-production of $B$ due to sphaleron processes must be dealt with.

In this work, motivated by its potential cosmological benefits, we consider the case of $|L_{\mu,\tau}| \sim \mathcal{O}(0.1-1)$ which in general requires even larger asymmetries  at high temperature, well above the electroweak scale.
Sphaleron processes in this case should be sufficinetly  suppressed in order to avoid baryon over-production.
%
%
Interestingly enough, it has been known for a long time that, if $|L = \sum_\alpha L_\alpha|$ is large enough, the electroweak symmetry can be never restored \cite{Linde:1976kh,Linde:1979pr} and the sphaleron processes can be exponentially suppressed, efficiently blocking the conversion of the total lepton number asymmetry into a baryon number asymmetry.
However, breaking the electroweak symmetry requires $|L|$ to be larger than about $10$ \cite{Bajc:1997ky} which is already too large to be consistent with CMB data \cite{Ade:2015xua}. Besides, a large enough suppression of the sphaleron processes requires even larger $|L|$. 
Hence, in order for $|L_\alpha| \sim \mathcal{O}(1)$ at the present universe to be consistent with observations, a much larger $|L|$ had to be generated above the electroweak scale temperature and diluted to a safe level well before BBN.

For all these reasons, we will discuss how one can have a fully consistent picture of a large lepton number asymmetry, including the high temperature lower bound on $|L|$ for symmetry-breaking, a large enough suppression of sphalerons, a plausible scenario of late-time mild entropy release, the generation of a large enough asymmetry, and a brief discussion about dark matter.

\section{Symmetry breaking due to a large total lepton number asymmetry}

\begin{figure*}[ht]
\begin{center}
\includegraphics[width=0.475\textwidth]{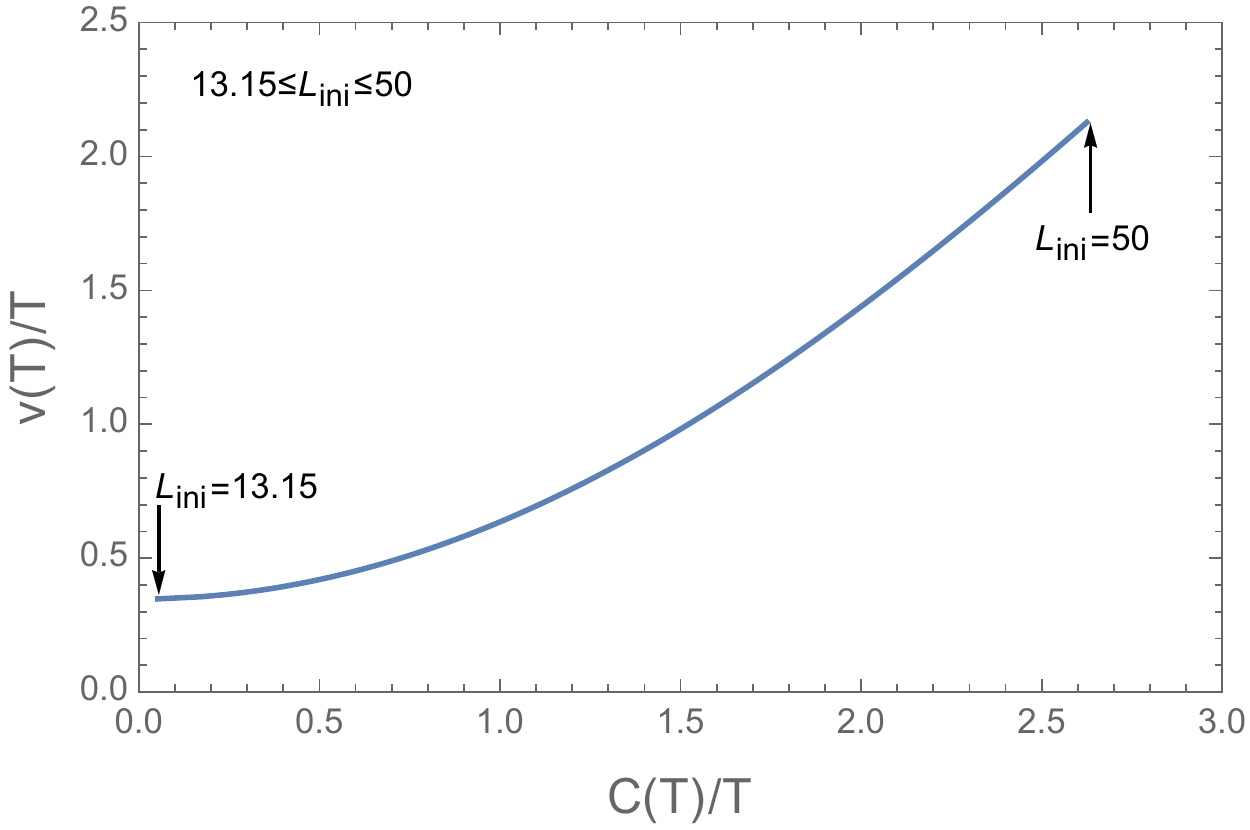}
\includegraphics[width=0.475\textwidth]{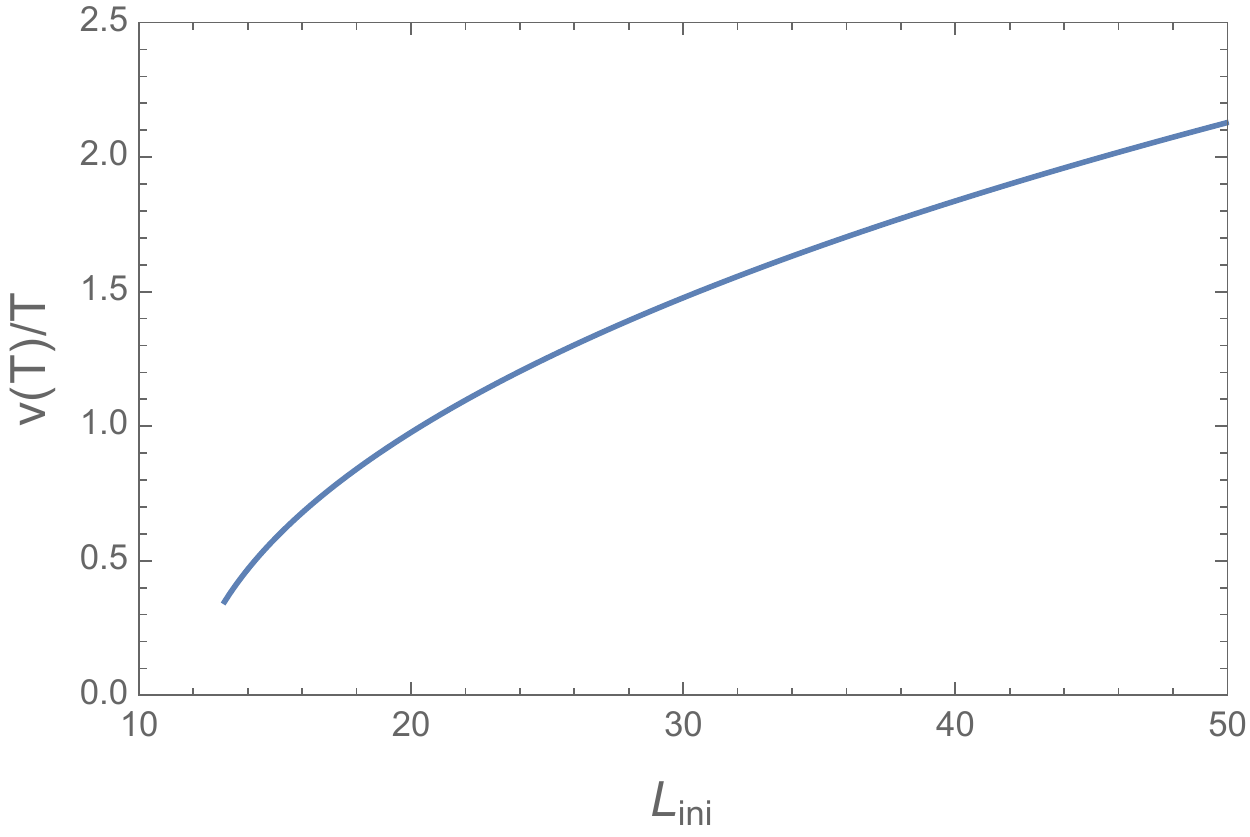}
\caption{\textit{Left}: Vacuum expectation values of CP-even Higgs ($v$) and gauge field ($C$) as a function of $L_{\rm ini}$. \textit{Right}: $v$ as a function of $L_{\rm ini}$. 
}
\label{fig:vevs-vs-L}
\end{center}
\end{figure*}
Conventionally, the electroweak symmetry breaking of the standard model (SM) is supposed to take place when the temperature of the Universe drops down to $T_{\rm ew} \sim 100 {\rm GeV}$.
However, it is known that the presence of a large net lepton number asymmetry (coming, for example,  from left-handed neutrinos) can cause a symmetry breaking of some SM gauge groups \cite{Linde:1976kh,Linde:1979pr}.
Specifically, in the standard model the vacuum expectation value (VEV) of the canonically normalized CP-even Higgs field is determined by a finite temperature effective potential of the form \cite{Bajc:1997ky}, 
\bea \label{Veff}
V_{\rm eff} &=& \frac{\lambda }{4} v^4 + \frac{\lambda'}{2}T^2 v^2 + \frac{g^2}{8} C^2 v^2 + \frac{n_L^2}{T^2}
\nonumber \\
&& + \frac{4 n_L^2 \l( 3 v^2 + 12 C^2 + 14 T^2 \r)}{54 C^2 v^2 + \l( 87 v^2 + 96 C^2 \r) T^2 + 112 T^4}
\eea
where $v$ is the VEV of the canonically normalized CP-even neutral Higgs field, $C$ the VEV of a gauge field, $n_L$ the net asymmetry of the lepton number density,
\beq
\lambda' = \frac{1}{12} \l[ 6 \lambda + y_\tau^2 + 3 y_t^2 + 3 y_b^2 + \frac{3}{4} \l( g'^2 + 3 g^2 \r) \r]
\eeq
assuming that the Yukawa coulings ($y_i$s) are dominated by the third generation contribution, 
and $g'$ is the gauge coupling of the $U(1)_Y$ gauge group of the SM. 
From \eq{Veff}, it is straightforward to see that $v(T)/T$ appears to be nonzero for $L_{\rm ini} \gtrsim 13.15$ with $L_{\rm ini}$ being the initial lepton number asymmetry at high energy, as shown in Fig.~\ref{fig:vevs-vs-L}.
It is clear then that an effective suppression of the sphaleron rate in order to avoid over-production of baryon number asymmetry requires even larger $L_{\rm ini}$. Such a large $L_{\rm ini}$ is difficult to be made consistent with CMB data unless there is enough amount of dilution caused by, for example, a late time entropy release taking place below $T_{\rm ew}$ but well before BBN.

%
%
%



\section{Entropy production from Peccei-Quinn sector}
A late-time entropy release can be obtained, for example, when the dominant energy content of the early universe is 
no longer given by radiation but by  non-relativistic particles which should decay before BBN.
In this case, the pre-existing lepton number asymmetry will be diluted by the entropy production 
associated to the decay.
Since we are interested in $L \sim \mathcal{O}(1)$ after the entropy release, the dilution factor should be 
\beq
\Delta = \frac{L_{\rm ini}}{L} \gtrsim \mathcal{O}(10)
\eeq
Now, the question is how we can realize such a mild late-time entropy generation.

MSSM flat-directions or stringy moduli are well motivated potential candidates for such a purpose.
However, generically, the former decays too early, and the latter produces too much entropy.  
On the other hand, the Peccei-Quinn field responsible for the axion solution of the strong CP problem \cite{Peccei:1977hh} may fit well to our purpose, since the energy density and decay rate of the saxion, the scalar partner of the axion, can be adjusted by independent parameters.
For example, assuming that the $U(1)$ Peccei-Quinn symmetry was broken before or during inflation, one can regard the axion coupling constant as a free parameter only lower bounded by astrophysical constraints \cite{Raffelt:2006cw}.

Depending on the specifics of the theoretical model and the cosmological scenarios, it is possible that saxion field $\phi$ can start coherent oscillations with respect to its true minimum when the expansion rate ($H$) of the universe becomes similar to its mass scale at zero-temperature.
The temperature at this epoch is 
\beq
T_{\rm osc} \approx \l( \frac{\pi^2}{90} g_*(T_{\rm osc}) \r)^{-1/4} \sqrt{m_\phi M_{\rm P}}
\eeq
where $g_*(T_{\rm osc})$ is the number of relativistic degrees of freedom, $m_\phi$ is the zero-temperature mass of saxion, and $M_{\rm P}$ is the reduced Planck mass.
The initial oscillation amplitude ($\phi_{\rm osc}$) is also model-dependent, but here we take it to be of the order of $\phi_0$, the zero-temperature VEV of saxion, for simplicity.
Then, saxion energy density starts dominating the universe when temperature drops to
\bea \label{Teq}
T_* &\approx& \frac{1}{6} \l( \frac{g_{*s}(T_{\rm osc})}{g_{*s}(T_*)} \r)^{1/3} \l( \frac{\phi_{\rm osc}}{M_{\rm P}} \r)^2 T_{\rm osc} 
\\
&\approx& c \l( \frac{\phi_{\rm osc}}{M_{\rm P}}  \r)^2 \l( \frac{M_{\rm P}}{m_\phi} \r)^{1/2} m_\phi 
\\
&\simeq& 27.4 {\rm GeV} \times  \l( \frac{\phi_{\rm osc}}{10^{14} {\rm GeV}} \r)^2 \l( \frac{m_\phi}{{\rm 10 TeV}} \r)^{1/2}
\eea
with
\beq
c \equiv \frac{1}{6} \l( \frac{g_{*s}(T_{\rm osc})}{g_{*s}(T_*)} \r)^{1/3} \l( \frac{\pi^2}{90} g_*(T_{\rm osc}) \r)^{-1/4} \sim \mathcal{O}(0.1)
\eeq
where $g_{*s}(T_{\rm osc})=200$ and $g_{*s}(T_*)=86.25$ were used for our numerical estimations.

The decay rate of the saxion depends on its couplings to SM particles.
In simple hadronic axion models \cite{Kim:1979if,Shifman:1979if}, saxions decay dominantly to light QCD-axions, 
bearing the potential  danger of forming a standard thermal background which could jeopardize  a successful BBN.
Hence, we  should have direct interactions of saxion to SM particles like, for example, in DFSZ axion scenarios \cite{Zhitnitsky:1980tq,Dine:1981rt}.
In this case, when kinematically allowed, the saxion can decay dominantly to the SM higgs  with a rate given by \cite{Kim:2008yu}
\beq
\Gamma_{s \to hh} = \frac{1}{2\pi} \frac{m_\phi^3}{\phi_0^2} \l( \frac{\mu}{m_\phi} \r)^4 \l( 1 - \frac{m_h^2}{m_\phi^2} \r)^{1/2}  
\eeq
where $\mu$ and $m_h=125.5 {\rm GeV}$ are the parameter of the Higgs bilinear in the minimal supersymmetric standard model (MSSM) and the observed mass of Higgs particle, respectively.  
Hence, for $\Gamma_s \simeq \Gamma_{s \to hh}$ the decay temperature is found to be
\bea \label{Td}
T_{\rm d} &=& \l( \frac{\pi^2}{90} g_*(T_{\rm d}) \r)^{-1/4} \sqrt{\Gamma_s M_{\rm P}}
\\
&\simeq& 2.86 {\rm GeV} \times \l( \frac{\mu}{m_\phi} \r)^2 \l( \frac{m_\phi}{10 {\rm TeV}} \r)^3 \l( \frac{10^{14} {\rm GeV}}{\phi_0} \r)^2 
\eea
where $g_*(T_{\rm d})=75.75$ and the instantaneous decay approximation were used.
From \eqs{Teq}{Td}, one finds
\bea \label{dilution}
\Delta &\approx& T_*/T_{\rm d}
\\
&\approx& 9.58 \l( \frac{m_\phi}{\mu} \r)^2 \l( \frac{10 {\rm TeV}}{m_\phi} \r)^{5/2} \l( \frac{\phi_0}{10^{14} {\rm GeV}} \r)^4
\eea
where we used $\phi_{\rm osc} = \phi_0$ in the second line.
Since we need only $\Delta = \mathcal{O}(10)$ with $T_{\rm d} < T_{\rm ew}$, we may consider the case of $T_* < T_{\rm ew}$ for simplicity.

\section{Lower bound of $L$ at high energy}
Lacking another source, it is natural to assume that the baryon number asymmetry is generated 
mainly from a large lepton number asymmetry via sphaleron processes.
In this case, the density of the baryon number asymmetry when sphaleron processes are practically terminated at $t_f$ (i.e., at $T \sim T_{\rm ew}$) is given by 
\beq \label{nB}
n_B \approx - \frac{1}{a^3} \int_{t_i}^{t_f} dt \Gamma_{\rm sph} a^3(t) n_L
\eeq
where we take $t_i$ to be the  time at which the generation of $L$ takes place, and ignore the back-reaction of $n_B$ to $n_L$ (i.e., the conversion of a tiny $n_B$ to $n_L$ via sphalerons ). 
The sphaleron rate after the  electroweak symmetry breaking is then given by \cite{Davidson:1994ni}
\beq
\Gamma_{\rm sph}(T) \sim \frac{10^{-2} m_W^7}{\alpha_{\rm W}^3 T^6} e^{-E_{\rm sph}/T}
\eeq
where $m_W = gv/2$ is the $W$-boson mass with $g$ and $v$ being the gauge coupling of the $SU(2)_L$ gauge symmetry 
of the SM and the VEV of the CP-even canonical Higgs field, respectively.  $\alpha_{\rm W} = g^2/4\pi$, and $E_{\rm sph} = \tilde{E} \times 2 m_W/\alpha_{\rm W}$ is the sphaleron energy with $\tilde{E} \approx 2$ for the standard model with a $125.5 {\rm GeV}$ Higgs mass \cite{Klinkhamer:1984di}.
Similarly, when the $SU(2)_L$ is broken at high temperature, well above the electroweak scale, we can take 
\beq \label{sph-rate}
\Gamma_{\rm sph}(T \gg T_{\rm ew}) \sim \frac{10^{-1} v(T)^7}{T^6} e^{-\frac{8 \pi v(T)}{g T}}
\eeq
where we keep $g$ being the low energy value \footnote{More accurately, $g$ should be temperature dependent, but we ignore such an effect since the change of $g$ will not be significant for the range of temperatures we explore on this work}.

It is important to stress that the matter-domination era starts at $T \ll T_{\rm ew}$ in the scenario we are considering.
Also, $v(T)/T \equiv r$ is nearly constant for $T \gg T_{\rm ew}$ in this set-up.
Therefore, as $T$ approaches $v(T=0)$, the conversion of $L$ to $B$ is practically shutted down.
Hence, regarding $r$ as a constant for $T \gtrsim T_f$, from \eq{nB} one finds
\beq
B(T_f) \sim - 10^{-1} \l( \frac{\pi^2}{90} g_*(T_f) \r)^{-1/2} \frac{M_{\rm P}}{T_f} r^7 e^{-\frac{8 \pi r}{g}}  L_{\rm ini}
\eeq
At late-times, including the dilution caused by the entropy release, the contribution of lepton number asymmetry to the baryon number asymmetry at present  becomes
\bea
B_0 &=& B(T_{\rm f} \gtrsim T_{\rm ew})/\Delta
\nonumber \\
&\sim& - 10^{-1} \l( \frac{\pi^2}{90} g_*(T_f) \r)^{-1/2} \frac{M_{\rm P}}{T_f} r^7 e^{-\frac{8 \pi r}{g}} L_0
\eea
where $L_0$ represents the net lepton number asymmetry at present.

\begin{figure}[t]
\begin{center}
\includegraphics[width=0.475\textwidth]{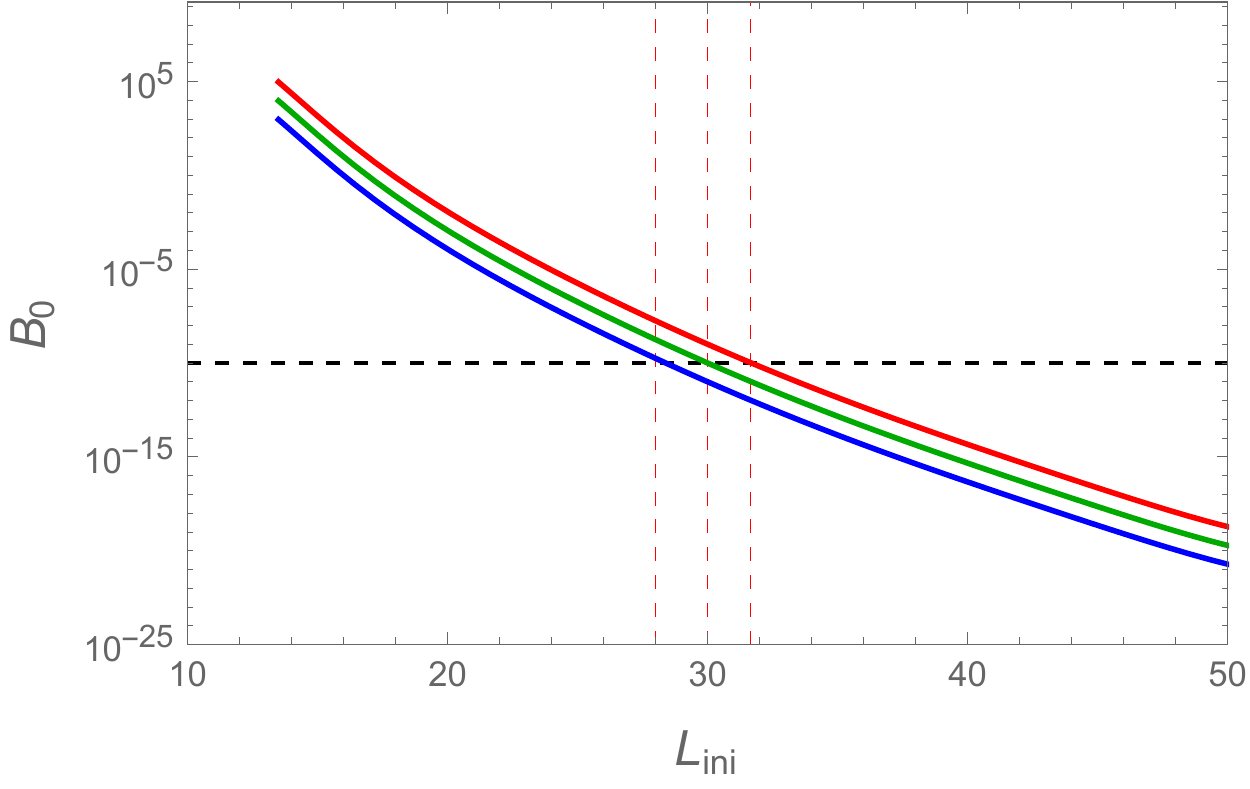}
\caption{Baryon number asymmetry at present as a function of the initial lepton number asymmetry at high energy for $T_f=v=246 {\rm GeV}$ and $L_0=10^{-2},0.1,1$ from left to right.
Dashed lines are indicating the value of $L_{\rm ini}$ for $B_0=10^{-10}$. 
}
\label{fig:B0-vs-L}
\end{center}
\end{figure}
In Fig.~\ref{fig:B0-vs-L}, we show the expected present baryon number asymmetry as a function of $L_{\rm ini}$ for $L_0 =10^{-2}, 0.1,$ and $ 1$.
From the figure, it can be clearly seen that $L_0 =1$ and  $B_0 \lesssim 10^{-10}$ requires 
\beq \label{Li-bnd}
L_{\rm ini} \gtrsim 32.4
\eeq 
which corresponds to $v(T)/T \gtrsim 1.57$.

\section{Leptogenesis}

The lower bound for the initial lepton number asymmetry in \eq{Li-bnd} is very large and therefore quite
challenging from the model building point of view.
However, such a large lepton number asymmetry can be generated in the early univrese, for example, by the 
Affleck-Dine mechanism \cite{Affleck:1984fy,Dine:1995uk,Dine:1995kz}.
Specifically, in Refs.~\cite{Casas:1997gx,McDonald:1999in}, it was shown that $L \sim \mathcal{O}(10)$ can be achieved either along the right-handed sneutrino direction or along the left-handed sneutrino field involved 
in the MSSM flat-directions.
Let us briefly describe the case of a MSSM flat-direction, following Ref.~\cite{McDonald:1999in}.

Let us consider a flat direction ($\Phi$) lifted by a $d=6$ superpotential term, $W = \lambda \Phi^6/M_{\rm P}^3$ with $\lambda$ being a numerical coupling constant.
In this case, through the Affleck-Dine mechanism, a large lepton number asymmetry can be generated when the expansion rate $H$ of the Universe becomes similar to the soft SUSY-breaking mass $m$ of the flat direction $\Phi$ around the origin.
We will assume that the Universe was dominated by radiation when the lepton number asymmetry is generated.
Then, at the generation of the asymmetry, the temperature of the Universe is given by 
\bea
T &\approx& \l( \frac{\pi^2}{90} g_*(T) \r)^{-1/4} \l( m M_{\rm P} \r)^{1/2}
\nonumber \\
&\simeq& 7.2 \times 10^{10} {\rm GeV} \l( \frac{g_*(T)}{200} \r)^{-1/4} \l( \frac{m}{10 {\rm TeV}} \r)^{1/2}
\eea
and the corresponding  number density of lepton asymmetry reads
\beq
n_L \approx m \phi_{\rm AD}^2 \sin \delta_{\rm CP}
\eeq
where $\phi_{\rm AD}$, the initial oscillation amplitude of Affleck-Dine field $\Phi$, is given by 
\beq
\phi_{\rm AD}^2 \approx \frac{2}{(5 \lambda^2)^{1/4}} \l( m M_{\rm P}^3 \r)^{1/2}
\eeq
and $\delta_{\rm CP}$ is the CP-violating phase.
Hence the lepton number asymmetry at high temperature is found to be
\bea \label{Li}
L_{\rm ini} &\approx& \frac{\pi^2}{\zeta(3) (5 \lambda^2)^{1/4}} \l( \frac{\pi^2}{90} g_*(T) \r)^{3/4} \sin \delta_{\rm CP} 
\nonumber \\
&\simeq& 56 \l( \frac{g_*(T)}{200} \r)^{3/4} \frac{\sin \delta_{\rm CP}}{\sqrt{\lambda}}
\eea
where we used that $H=m$. 
It becomes apparent then that this scenario can easily  
satisfy the condition in \eq{Li-bnd} for plausible ranges of $\delta_{\rm CP}$ and $\lambda$.

It should be noted that, when the lepton number asymmetry is generated, for $m \sim {\rm TeV}$ the temperature of the Universe is very high and gravitinos are likely to be over-produced if the mass of the gravitino is similar to the scale of the soft SUSY breaking masses of scalar fields.
Hence, in this case, in the presence of only mild entropy production due to saxion decays it is necessary to have gravitinos much heavier than ${\rm TeV}$ scale, otherwise late time decays of gravitinos would prevent a successful BBN.
This is the reason we consider $\mathcal{O}(10) {\rm TeV}$ as the typical mass scales of $m_\phi$ and $\mu$ in 
the earlier discussion.
%
%
Such a scale can arise naturally in the \textit{pure gravity mediation} scenario \cite{Ibe:2011aa}.
In such a scheme, the gravitino problem can be avoided and a right amount of relic dark matter at present can be 
provided by the  wino playing the role of the  lightest supersymmetric particle.
However, due to the late-time entropy release in our scenario, wino dark matter would be sub-dominant in the end.
The main component of dark matter can come from the axion misalignment.
Since the Peccei-Quinn symmetry is assumed to be broken before or during inflation, the misalignment is determined simply by the initial phase shift of the Peccei-Quinn field, and can be freely chosen to provide the  
right amount of axion dark matter.

\section{Conclusions}
In this paper, we discuss a complete scenario able to accomodate a large lepton number asymmetry surviving up 
to this date.

A good fit to astrophysical and cosmological data may require $\mathcal{O}(1)$ lepton number asymmetries $L_\alpha$ (normalized by photon number density) of neutrino flavors $\nu_\alpha$ ($\alpha = \mu, \tau$ only).
In view of the standard model (or its simple supersymmetric extensions), such large lepton number asymmetries are dangerous since they could end up generating too much baryon number asymmetry due to anomalous electroweak process (called sphalerons), unless the net asymmetry is comparable to or smaller than the observed amount of baryon number asymmetry (i.e., $B \sim 10^{-10}$) or the asymmetry is generated well after the electroweak phase transition.

In the present case however we are interested in  an early generation of $L_\alpha$s, i.e. 
 well before the electroweak phase transition. In such a case even if an extreme fine tuning to get $L \equiv \sum_\alpha L_\alpha \lesssim B$ from $L_\alpha \sim \mathcal{O}(1)$ may be possible, it is doomed to be inconsistent with BBN bound on the lepton number asymmetry of electron neutrinos  due to neutrino oscillations \cite{Lunardini:2000fy,Dolgov:2002ab,Wong:2002fa,Abazajian:2002qx,Mangano:2010ei,Mangano:2011ip,Castorina:2012md,Johns:2016enc}.
On the other hand, interestingly enough, sphaleron processes 
can be suppressed if there is a large enough net lepton number asymmetry which can cause electroweak symmetry-breaking at temperature much higher than electroweak scale, avoiding this way  the danger of baryon over-production.

In this work, we show that for such an idea of large suppression of sphaleron to work,  
a $|L| \sim \mathcal{O}(1)$ at present requires the net generated asymmetry to be
\beq
|L|_{\rm ini} \gtrsim 32
\eeq 
at its original birth scale, a high energy scale well above the electroweak temperature.
This implies that there has to be some amount of dilution of $L_{\rm ini}$  attenuating  it  
up to $|L|\sim \mathcal{O}(1)$ at present.
We show that such a  mild dilution can be easily obtained by the late time decay of the saxion, 
the scalar partner of the Peccei-Quinn axion.
The required $L_{\rm ini}$ can be generated naturally by the 
Affleck-Dine mechanism with a superpotential term $W = \lambda \Phi^6/M_{\rm P}^3$ for a MSSM flat direction $\Phi$.  
In order to avoid the danger of gravitino over-production, the mass of the 
gravitino was assumed to be comparable to or larger than $\mathcal{O}(10) {\rm TeV}$, and so was the mass of the saxion.
As a result, the axion coupling constant was constrained to be comparable to or larger 
than $\mathcal{O}(10^{14}) {\rm GeV}$.
Dark matter is expected to be mostly from axion misalignment under the assumption that the breaking of Peccei-Quinn symmetry took place before or during primordial inflation.

\section{Acknowledgement} 
GB acknowledges support from the MEC and FEDER (EC) Grants SEV-2014-0398 and FPA2014-54459 and the Generalitat Valenciana under grant PROMETEOII/2013/017. 
This project has received funding from the European Union's Horizon 2020 research and innovation program under the Marie Sklodowska-Curie grant Elusives ITN agreement No 674896  and InvisiblesPlus RISE, agreement No 690575. 
It was also supported by research funds of Chonbuk National University in 2017.

\end{document}